\documentstyle[preprint,aps]{revtex}

\begin{document}
\draft
\title{{\bf Renormalization of Quantum Anosov Maps:}\\
{\bf Reduction to Fixed Boundary Conditions}}
\author{{\bf Itzhack Dana}}
\address{Minerva Center and Department of Physics, Bar-Ilan University,
Ramat-Gan 52900, Israel}
\maketitle

\begin{abstract}
A renormalization scheme is introduced to study quantum Anosov maps (QAMs)
on a torus for general boundary conditions (BCs), whose number ($k$) is
always finite. It is shown that the quasienergy eigenvalue problem of a
QAM for {\em all} $k$ BCs is exactly equivalent to that of the
renormalized QAM (with Planck's constant $\hbar ^{\prime }=\hbar /k$) at
some {\em fixed} BCs that can be of four types. The quantum cat maps are,
up to time reversal, fixed points of the renormalization transformation.
Several results at fixed BCs, in particular the existence of a complete
basis of ``crystalline'' eigenstates in a classical limit, can then be
derived and understood in a simple and transparent way in the general-BCs
framework.\newline
\end{abstract}

\pacs{PACS numbers: 05.45.Mt, 03.65.Ca, 03.65.Sq, 11.10.Gh}

Nonintegrable systems whose dynamics can be reduced to a 2D torus in
phase space have attracted much attention in the quantum-chaos
literature. When quantizing such a system on the torus, the admissible
quantum states must satisfy proper boundary conditions (BCs), i.e.,
they have to be periodic on the torus up to constant phase factors
specified by a Bloch wave vector ${\bf w}$. If the Hamiltonian of the
system is periodic in phase space, e.g., the kicked Harper model
\cite{l,d1,d2,d3}, its classical dynamics can be reduced to the toral
phase space of one unit cell of periodicity, and all Bloch wave
vectors ${\bf w}$ in some Brillouin zone (BZ) are allowed. The
sensitivity of the eigenstates to continuous variation of ${\bf w}$ in
the BZ may be characterized topologically by a Chern index
\cite{l,d1,d2,d3} and appears to be strong for eigenstates spread over
the chaotic region and weak for eigenstates localized on stability
islands.\newline

In general, however, the Hamiltonian of a system whose dynamics can be
reduced to a torus is {\em not} periodic in phase space. An extreme
case is that of the purely chaotic, Anosov ``cat maps''
\cite{va,hb,pv,d4,ke,k,de,egi,bb,n}, whose Hamiltonians are quadratic
in the phase-space variables \cite{ke}. When quantizing these systems,
it turns out that only a {\em finite} set of ${\bf w}$'s in the BZ is
allowed \cite{k,de,egi,bb,n,kmr}, see Eq. (\ref{cw}) below, and this
set {\em increases} with increasing chaotic instability. However, due
to the lack of 2D continuity in this set, the sensitivity to
variations in ${\bf w}$ cannot be characterized by a Chern index
\cite{kmr}. A large class of cat maps was first quantized by Hannay
and Berry \cite{hb} for ${\bf w=0}$ (strict periodicity on the torus)
and general BCs were considered in subsequent works
\cite{k,de,egi,bb,n,kmr}. An important recent work \cite{n} has shown
the existence of eigenstates whose Husimi density is periodic on a
lattice in the torus and whose Husimi zeros form a crystal, for all
Planck's constants in a sequence tending to zero. These
``crystalline'' eigenstates are the first example of exactly
equidistributed eigenstates of quantized chaotic systems in a
classical limit.\newline

An atypical feature of quantum cat maps is their high spectral
degeneracy, which increases in the classical limit
\cite{hb,ke}. Typical spectral properties, fitting generic eigenvalue
statistics, are already found by quantizing torus maps that are very
slight perturbations of the cat maps \cite{pc,pc1,pc2}. According to
Anosov's theorem \cite{va}, these maps have essentially the same
classical dynamics, in particular they are purely chaotic, as the
unperturbed cat maps, which are structurally stable. This ceases to be
the case for larger perturbations that cause bifurcations generating
elliptic islands and thus a mixed phase space \cite{pc2}. However, the
quantum BCs for a perturbed cat map are the same as those of the
unperturbed cat map, independently of the size of the perturbation
\cite {kmr}, see Eq. (\ref{cw}) below. Because of this reason and to
simplify terms, general perturbed cat maps are referred to as Anosov
maps in this Letter. The importance of these maps is in that they may
be viewed as {\em generic} torus maps on the basis of a general
expression for a smooth torus map derived recently \cite{kmr}, see
below.\newline

In this Letter, we introduce a renormalization scheme to study quantum
Anosov maps (QAMs) for general BCs. We show that the eigenvalue
problem of a QAM for {\em all} BCs, determining the {\em full} quantum
dynamics, can be exactly reduced to a problem at {\em fixed} BCs. More
specifically, consider a QAM given by the evolution operator $\hat{U}$
quantizing a classical Anosov map. Quantization on a torus requires a
Planck's constant $\hbar $ to satisfy $2\pi /\hbar =p$, an
integer. The finite number of BCs is denoted by $k$, which depends
only on the classical unperturbed cat map. We define a renormalization
transformation ${\cal R}$ of $\hat{U}$ generating a QAM
$\hat{U}^{\prime}={\cal R}(\hat{U})$ on the same torus. The number of
BCs for $\hat{U}^{\prime }$ is also $k$ and $\hat{U}^{\prime }$ is
associated with a renormalized Planck's constant $\hbar ^{\prime
}=\hbar /k$. We then show that the quasienergy eigenvalue problem for
$\hat{U}$ for {\em all} $k$ BCs is equivalent, by a unitary
transformation accompanied by a scaling of variables, to that for
$\hat{U}^{\prime }$ at some {\em fixed} BCs. The latter can be of four
types, corresponding to the four ``symmetry'' points in the
renormalized BZ (see Table 1 below). The quantum cat maps are always
fixed points of ${\cal R}$ in the reflection-hyperbolic case. In the
ordinary-hyperbolic case, they are fixed points of ${\cal R}$ only if
some conditions are satisfied. Otherwise, they are fixed points of
${\cal R}$ accompanied by time reversal. Thus, for general $\hat{U}$,
the QAMs $\hat{U} ^{(n)}={\cal R}^{n}(\hat{U})$ represent
perturbations of a given quantum cat map, or its time inverse, in a
classical limit, $\hbar ^{(n)}=\hbar /k^{n}\rightarrow 0$. The
spectrum and eigenstates of $\hat{U}$, $\hat{U} ^{\prime }$,...,
$\hat{U}^{(n-1)}$ for all $k$ BCs and $n>1$ can be fully reproduced
from those of $\hat{U}^{(n)}$ at fixed BCs corresponding to either
strict {\em periodicity} for $kp$ even or {\em antiperiodicity} for
$kp$ odd. As a first application of all this scheme, we show that some
results at fixed BCs can be derived and understood in a simple and
transparent way in the general-BCs framework. In particular, it is
easy to show in this framework that the eigenstates of $\hat{U}^{(n)}$
at the fixed BCs are all ``crystalline''. Unlike general crystalline
eigenstates of the quantum cat maps \cite{n}, the crystal features
persist for arbitrarily large perturbations.\newline

We denote by ${\bf z}=(u,\ v)$ the phase-space variables, $[\hat{u},\
\hat{v}]=i\hbar $, and we assume that the classical dynamics can be
reduced to a $2\pi \times 2\pi$ torus $T^{2}$, where it is described
by an Anosov map $M$. In general, a smooth map $M$ on $T^{2}$ can be
written as ${\bf z} \rightarrow M({\bf z})=\overline{M}({\bf z})$
mod$\ 2\pi $, where the ``lifted'' map $\overline{M}$, defined on the
entire phase plane $(u,\ v)$, can be expressed uniquely as the
composition of two maps, $\overline{M} =M_{A}\circ M_{1}$
\cite{kmr}. Here $M_{A}$ is a linear map, $M_{A}({\bf z)} =A\cdot {\bf
z}$, where $A$ is a $2\times 2$ integer matrix with $\det (A)=1$; by
``Anosov'' we just mean that $|$Tr$(A)|>2$, a condition generically
satisfied by $A$. The map $M_{1}$ is defined by $M_{1}({\bf z})={\bf
z}+{\bf F}({\bf z})$, where ${\bf F}({\bf z})$ is a $2\pi $-periodic
vector function of ${\bf z}$. For ${\bf F}({\bf z})={\bf 0}$, $M$ is a
``cat map''. The quantization of $\overline{M}$ is the unitary
operator $\hat{U}=\hat{U}_{A} \hat{U}_{1}$, where $\hat{U}_{A}$ and
$\hat{U}_{1}$ are the quantizations of $M_{A}$ and $M_{1}$,
respectively \cite{kmr}. In the $u$ representation,
\begin{equation}
\langle u_{2}|\hat{U}_{A}|u_{1}\rangle _{\hbar }=\left( \frac{1}{2\pi
i\hbar A_{1,2}}\right) ^{1/2}\exp \left[ \frac{i}{2\hbar
A_{1,2}}\left( A_{1,1}u_{1}^{2}-2u_{1}u_{2}+A_{2,2}u_{2}^{2}\right)
\right] .  \label{UA}
\end{equation}
We shall assume that $M_{1}$ is the map for a Hamiltonian which is
periodic in phase space with unit cell $T^{2}$. As shown in
Ref. \cite{kmr}, this is the case if and only if $\int_{T^{2}}{\bf
F}({\bf z})\,d{\bf z=0}$. Then $\hat{U}_{1}$ is the one-step evolution
operator for the Weyl quantization of this Hamiltonian and is a
periodic operator function $\hat{U}_{1}({\bf \hat{z}};\ \hbar )$,
representable by a well defined Fourier expansion.  Quantization on a
torus requires now that $\hbar =2\pi /p$, $p$ integer. The QAM
quantizing $M$ is $\hat{U}$ restricted to act on the simultaneous
eigenstates of the commuting phase-space translations on $T^{2}$,
$\hat{D}_{1}=\exp (ip\hat{u})$ and $\hat{D}_{2}=\exp (-ip\hat{v})$;
the corresponding eigenvalues are $\exp (ipw_{1})$ and $\exp
(-ipw_{2})$, where $(w_{1},\ w_{2})={\bf w}$ is the Bloch wave vector
specifying the toral BCs \cite{d1}. An eigenstate $\Psi _{{\bf w}}$ of
$\hat{D}_{1}$ and $\hat{D}_{2}$ can be an eigenstate of $\hat{U}$ only
for those values of ${\bf w}$ in the Brillouin zone (BZ: $0\leq
w_{1},\ w_{2}<2\pi /p$ ) satisfying the equation
\cite{k,de,egi,bb,n,kmr}
\begin{equation}
A\cdot {\bf w}={\bf w}+\pi {\bf y\ }\text{mod }2\pi /p, \label{cw}
\end{equation}
where ${\bf y}\equiv (A_{1,1}A_{1,2},\ A_{2,1}A_{2,2})$. We write the
general solution of Eq. (\ref{cw}) as follows:
\begin{equation}
{\bf w}=(2\pi /p)B\cdot ({\bf r}+pE^{-1}\cdot {\bf y}/2)\text{ mod
}2\pi /p,
\label{sw}
\end{equation}
where $B=(A-I)^{-1}E$, $I$ is the identity matrix, $E$ is an arbitrary
integer matrix with $\det (E)=\pm 1$, and ${\bf r}$ is an integer vector
labeling the solutions. There are precisely $k=|\det
(B^{-1})|=|2-$Tr$(A)|$ distinct vectors (\ref{sw}), as the number of fixed
points of $M_{A}$ mod $2\pi$ \cite {ke}, forming a {\em lattice} in the
BZ. We denote by ${\cal S}$ the space of states $\Psi _{{\bf w}}$\ for all
these $k$ values of ${\bf w}$. The subspace ${\cal S}_{{\bf w}}$ of ${\cal
S}$ with a fixed value of ${\bf w}$ is $p$-dimensional, i.e., it is
spanned by a basis of $p$ independent states \cite{d1,d3}, whose general
expression in the $u$ representation is \cite {note}
\begin{equation}
\Psi _{b,{\bf w}}(u)=\sum_{m=0}^{p-1}\phi _{b}(m;\ {\bf
w})\sum_{l=-\infty }^{\infty }e^{ilpw_{2}}\delta (u-w_{1}-2\pi
m/p-2\pi l), \label{qes}
\end{equation}
where $b=1,...,\ p$. Such a basis is formed, naturally, by the $p$
eigenstates of $\hat{U}$ at fixed ${\bf w}$.\newline

We now introduce the torus $T_{B}^{2}$, defined by the vectors ${\bf
R} _{j}=2\pi k(B_{1,j},\ B_{2,j})$, $j=1,\ 2$; $kB$ has integer
entries and $T_{B}^{2}$ contains precisely $k$ tori $T^{2}$. Since
$B^{-1}AB=E^{-1}AE$ is an integer matrix, the superlattice with unit
cell $T_{B}^{2}$ is invariant under $A$, so that the map
$\overline{M}$ modulo $T_{B}^{2}$, denoted by $M^{(B)}$, is well
defined. To continue, we shall first work out in detail the
reflection-hyperbolic case of Tr$(A)<-2$, choosing $E=I$, so that
$[A,\ B]=0$ and $\det (B)>0$. We shall then specify the changes to be
made in the ordinary-hyperbolic case of Tr$(A)>2$. Let us perform the
linear transformation of variables
\begin{equation}
{\bf z}=kB\cdot {\bf z}^{\prime }=\sqrt{k}C\cdot {\bf z}^{\prime },
\label{rt}
\end{equation}
where $C=\sqrt{k}B$ and ${\bf z}^{\prime }=(u^{\prime },\ v^{\prime
})$. Eq.  (\ref{rt}) is the combination of a linear canonical
transformation [$\det (C)=+1$, since $\det (B)>0$] with a scaling by a
factor $\sqrt{k}$. Using $[A,\ B]=0$, it is easy to check that the map
$M^{(B)}$ above is transformed by (\ref{rt}) into a map $M^{\prime }$
on $T^{2}$ in the ${\bf z}^{\prime }$ variables, $M^{\prime }({\bf
z}^{\prime })=$ $\overline{M}^{\prime }({\bf z} ^{\prime })$ mod $2\pi
$, with $\overline{M}^{\prime }=M_{A}^{\prime }\circ M_{1}^{\prime }$
. Here $M_{A}^{\prime }({\bf z}^{\prime })=A\cdot {\bf z} ^{\prime }$
and $M_{1}^{\prime }({\bf z}^{\prime })=(kB)^{-1}\cdot M_{1}({\bf
z}=kB\cdot {\bf z}^{\prime})$.  The renormalization transformation
${\cal R}_{c}$ in the classical case is then defined by $M^{\prime
}({\bf z} ^{\prime })={\cal R}_{c}[M({\bf z})]$ \cite{rm}. Clearly,
the cat maps, with ${\bf F}({\bf z})={\bf 0}$, are fixed points of
${\cal R}_{c}$, i.e., $M^{\prime }({\bf z}^{\prime }={\bf z})=M({\bf
z})$.\newline

The quantum version of (\ref{rt}) implies that $[\hat{u}^{\prime },\
\hat{v} ^{\prime }]=i\hbar ^{\prime }$, where $\hbar ^{\prime }=\hbar
/k=2\pi /p^{\prime }$, $p^{\prime }\equiv kp$. The quantization
$\hat{U}^{\prime }$ of $M^{\prime }({\bf z}^{\prime })$ is simply
$\hat{U}$ expressed in terms of $({\bf \hat{z}}^{\prime },\ \hbar
^{\prime })$ and acting on the space of\ the simultaneous eigenstates
$\Psi _{{\bf w}^{\prime }}^{\prime }$ of the phase-space translations
on $T^{2}$ in the ${\bf z}^{\prime }$ variables, $\hat{D}_{1}^{\prime
}=\exp (ip^{\prime }\hat{u}^{\prime })$ and $\hat{D}_{2}^{\prime
}=\exp (-ip^{\prime }\hat{v}^{\prime })$. It is easy to show that
\begin{equation}
\hat{D}_{j+1}^{\prime }=\hat{D}({\bf
R}_{j})=(-1)^{pk^{2}B_{1,j}B_{2,j}}\hat{
D}_{1}^{kB_{2,j}}\hat{D}_{2}^{kB_{1,j}} \label{WH}
\end{equation}
($j=1,\ 2$, $\hat{D}_{3}^{\prime }\equiv \hat{D}_{1}^{\prime }$),
where $\hat{D}({\bf R}_{j})$ are precisely the Weyl-Heisenberg
phase-space translations on $T_{B}^{2}$. By expressing
$\hat{U}=\hat{U}_{A}\hat{U}_{1}$ in terms of $({\bf \hat{z}}^{\prime
},\ \hbar ^{\prime })$, using also the theory of linear quantum
canonical transformations \cite{mq}, we obtain the expected result
$\hat{U}^{\prime }=\hat{U}_{A}^{\prime }\hat{U}_{1}^{\prime } $. Here
the $u^{\prime }$ representation of \ $\hat{U}_{A}^{\prime }$ is given
by (\ref{UA}) with $u$ and $\hbar $ replaced by $u^{\prime }$ and
$\hbar ^{\prime }$, respectively, and $\hat{U}_{1}^{\prime }$ is the
operator function $\hat{U}_{1}^{\prime }({\bf \hat{z}}^{\prime };\
\hbar ^{\prime })= \hat{U}_{1}({\bf \hat{z}}=kB\cdot {\bf
\hat{z}}^{\prime };\ \hbar =k\hbar ^{\prime })$ [the function
$\hat{U}_{1}({\bf \hat{z}};\ \hbar )$ was defined above]. The
renormalization transformation ${\cal R}$ is then defined by
$\hat{U}^{\prime }={\cal R}(\hat{U})$. The quantum cat maps
($\hat{U}=\hat{U}_{A}$) are fixed points of ${\cal R}$, i.e., $\langle
u_{2}^{\prime }=u_{2}| \hat{U}^{\prime }|u_{1}^{\prime }=u_{1}\rangle
_{\hbar ^{\prime }=\hbar }=\langle u_{2}|\hat{U}|u_{1}\rangle _{\hbar
}$.\newline

The space of states $\Psi _{{\bf w}^{\prime }}^{\prime }$ for all the
$k$ allowed values of ${\bf w}^{\prime }$ will be denoted by ${\cal
S}^{\prime }$. We now show that ${\cal S}$ coincides with the subspace
${\cal S}_{{\bf w} _{0}^{\prime }}^{\prime }$ of ${\cal S}^{\prime }$
associated with a particular value ${\bf w}_{0}^{\prime }$. Thus,
${\cal S}^{\prime }$ is $k$ times larger than ${\cal S}$. To show
this, let us apply $\hat{D} _{j}^{\prime }$, $j=1,\ 2$, on a state
$\Psi _{{\bf w}}$ of ${\cal S}$.  Using (\ref{sw}), (\ref{WH}), and
the fact that $\hat{D}_{j}\Psi _{{\bf w}}=$ $\exp
[i(-1)^{j+1}pw_{j}]\Psi _{{\bf w}}$, $j=1,\ 2$, we obtain
\begin{equation}
\hat{D}_{j}^{\prime }\Psi _{{\bf w}}=(-1)^{pA_{j,j+1}}\Psi _{{\bf w}}
\label{Dpw}
\end{equation}
($A_{2,3}\equiv A_{2,1}$). Rel. (\ref{Dpw}) means that {\em all}
$\Psi_{{\bf w}}$ in ${\cal S}$ are eigenstates of $\hat{D}_{j}^{\prime
}$, $j=1,\ 2$, associated with the {\em same} renormalized Bloch wave
vector ${\bf w} _{0}^{\prime }$. The latter can assume only four
values, depending on the matrix $A$, see Table 1.
\[
\begin{tabular}{||c||c||c||}
\hline\hline & $k$ even & $k$ odd \\ \hline\hline $p$ even & ${\bf
w}_{0}^{\prime }={\bf 0}$ & ${\bf w}_{0}^{\prime }={\bf 0}$ \\
\hline\hline $p$ odd & ${\bf w}_{0}^{\prime }=\left( \frac{A_{1,2}\pi
}{p^{\prime }},\ \frac{A_{2,1}\pi }{p^{\prime }}\right) $ mod
$\frac{2\pi }{p^{\prime }}$ & $ {\bf w}_{0}^{\prime }=\left( \frac{\pi
}{p^{\prime }},\ \frac{\pi } {p^{\prime }}\right) $ \\ \hline\hline
\end{tabular}
\]
\[
\text{Table 1.}
\]
It is easy to show that ${\bf w}_{0}^{\prime }$ is indeed an allowed
value of ${\bf w}^{\prime }$ in all four cases. Thus, ${\cal S}_{{\bf
w} _{0}^{\prime }}^{\prime }$ includes ${\cal S}$, but since both
${\cal S}_ {{\bf w}_{0}^{\prime }}^{\prime }$ and ${\cal S}$ are
$kp$-dimensional, they coincide. This completes the proof. Now, by the
definition above of $\hat{U} ^{\prime }$, the restriction of
$\hat{U}^{\prime }$ to ${\cal S}_{{\bf w} _{0}^{\prime }}^{\prime
}={\cal S}$ is just $\hat{U}$. The $kp$ eigenstates of $\hat{U}$ for
all $k$ BCs are then precisely the $p^{\prime }$ eigenstates of
$\hat{U}^{\prime }$ associated with the value of ${\bf w} _{0}^{\prime
}$ in Table 1. When referred to $\hat{U}^{\prime }$, however, these
eigenstates should be expressed in a representation based on the
operator ${\bf \hat{z}}^{\prime}$. If the $kp$ eigenstates of
$\hat{U}$ are $\Psi _{b,{\bf w}}(u)$ in the $u$ representation, see
(\ref{qes}), their $u^{\prime}$ representation will be obtained by
applying to $\Psi_{b,{\bf w}}(u)$ the unitary transformation
corresponding to a linear canonical transformation \cite{mq} with
matrix $C$, after scaling $u^{\prime }$ by a factor $\sqrt{k}$. The
eigenstates of $\hat{U}^{\prime }$ for ${\bf w} ^{\prime }={\bf
w}_{0}^{\prime }$ are thus given by
\begin{equation}
\Psi _{b^{\prime },{\bf w}_{0}^{\prime }}^{\prime }(u^{\prime
})=\left( \frac{p}{4\pi ^{2}B_{1,2}}\right) ^{1/2}\int\limits_{-\infty
}^{\infty }du\exp \left[ \frac{-ip}{4\pi B_{1,2}}\left(
kB_{1,1}u^{\prime }{}^{2}-2u^{\prime }u+B_{2,2}u^{2}\right) \right]
\Psi _{b,{\bf w}}(u),
\label{qep}
\end{equation}
where $b^{\prime }=b^{\prime }(b,\ {\bf w})$ takes precisely all its
$p^{\prime }$ values when $b$ and ${\bf w}$ take all their $p$ and $k$
values, respectively; conversely, $\Psi _{b,{\bf w}}(u)$ can be fully
reproduced from $\Psi _{b^{\prime },{\bf w}_{0}^{\prime }}^{\prime
}(u^{\prime })$ by inverting Rel. (\ref{qep}) and determining ${\bf
w}$ by applying $\hat{D}_{1}$ and $\hat{D}_{2}$ on $\Psi _{b,{\bf
w}}(u)$. If the quasienergies of $\hat{U}$ are $\omega _{b}({\bf w})$,
those of $\hat{U} ^{\prime }$ for ${\bf w}^{\prime }={\bf
w}_{0}^{\prime }$ are $\omega _{b^{\prime }(b,{\bf w})}^{\prime }({\bf
w}_{0}^{\prime })=\omega_{b}({\bf w })$. The latter relation and
Rel. (\ref{qep}) show the equivalence between the quasienergy
eigenvalue problem for $\hat{U}$ for all $k$ BCs and that for
$\hat{U}^{\prime }$ at the fixed BCs given by ${\bf w}^{\prime }={\bf
w} _{0}^{\prime }$.\newline

The case of Tr$(A)>2$ can be treated similarly only if the integer
matrix $E$ in $B=(A-I)^{-1}E$ can be chosen so that $[A,\ E]=0$ and
$\det (E)=-1$. Then one has again $[A,\ B]=0$ and $\det (C)=+1$ in
(\ref{rt}), leading to the same results as above. If $A=K^{2l}$, where
$K$ is any integer matrix with $\det (K)=-1$ and $l$ is an integer,
one can choose $E=K$. In general, an integer matrix $E$ having the
properties above does not exist, and we then make the simple choice
$E_{1,1}=-E_{2,2}=1$, $E_{1,2}=E_{2,1}=0$, corresponding to time
reversal. As a result, in all the expressions and equations involving
the renormalized quantities, including in Table 1, $A$ is replaced by
$A^{\prime }=E^{-1}AE$. Thus, the quantum cat maps are now fixed
points of ${\cal R}$ accompanied by time reversal: Given the
eigenstates $\Psi _{b,{\bf w}}(u)$ and quasienergies $\omega _{b}({\bf
w})$ of $\hat{U}_{A}$ for all BCs, its eigenstates and quasienergies
for $\hbar ^{\prime }=2\pi /p^{\prime }$ and ${\bf w}^{\prime }={\bf
w}_{0}^{\prime }$ are, respectively, $\Psi _{b^{\prime },{\bf
w}_{0}^{\prime }}^{\prime \ast }(u^{\prime })$ and $\omega _{b^{\prime
}(b,{\bf w})}^{\prime }({\bf w} _{0}^{\prime })=-\omega _{b}({\bf
w})$, where $\Psi _{b^{\prime },{\bf w} _{0}^{\prime }}^{\prime
}(u^{\prime })$ is given by (\ref{qep}).  Time-reversal symmetry
($A^{\prime }=A^{-1}$) allows to choose $\Psi _{b^{\prime },{\bf
w}_{0}^{\prime }}^{\prime }(u^{\prime })$ as real, $\Psi _{b^{\prime
},{\bf w}_{0}^{\prime }}^{\prime \ast }(u^{\prime })=\Psi _{b^{\prime
},{\bf w}_{0}^{\prime }}^{\prime }(u^{\prime })$.\newline

By iterating ${\cal R}$, one obtains a sequence of QAMs
$\hat{U}^{(n)}={\cal R}^{n}(\hat{U})$ on $T^{2}$, associated with the
Planck's constants $\hbar ^{(n)}=2\pi /p^{(n)}$, $p^{(n)}\equiv
k^{n}p$. General $\hat{U}^{(n)}=\hat{U} _{A}^{(n)}\hat{U}_{1}^{(n)}$
represent perturbations of the quantum cat map $\hat{U}_{A}$, or its
time inverse $\hat{U}_{A^{\prime }}$ (only for $n$ odd), in a
classical limit, $n\rightarrow \infty $. The perturbation $\hat{U}
_{1}^{(n)}({\bf \hat{z}}^{(n)};\ \hbar ^{(n)})$ is periodic in ${\bf
\hat{z}} ^{(n)}$ with a unit cell $T_{n}^{2}=(kB)^{-n}\cdot T^{2}$,
which is $k^{n}$ times smaller than $T^{2}$. In the generalization of
Table 1 to $n>1$, $p^{(n-1)}=k^{n-1}p$ is always even when $k$ is
even, so that ${\bf w} _{0}^{(n)}$ can take {\em only two} values:
${\bf w}_{0}^{(n)}={\bf 0}$ (for $kp$ even) and ${\bf
w}_{0}^{(n)}=(\pi /p^{(n)},\ \pi /p^{(n)})$ (for $kp$ odd),
corresponding to strictly periodic and antiperiodic BCs, respectively.
The eigenstates of $\hat{U}^{(n)}$ for ${\bf w}^{(n)}={\bf
w}_{0}^{(n)}$ are connected with those of $\hat{U}^{(n-1)}$ for all
$k$ BCs by a relation analogous to Rel. (\ref{qep}). The quasienergies
are related by $\omega _{b^{(n)}}^{(n)}({\bf w}_{0}^{(n)})=\omega
_{b^{(n-1)}}^{(n-1)}({\bf w} ^{(n-1)})$. Thus, the spectrum and
eigenstates of $\hat{U},...,\ \hat{U} ^{(n-1)}$ for all BCs can be
fully reproduced from those of $\hat{U}^{(n)}$ for ${\bf w}^{(n)}={\bf
w}_{0}^{(n)}$.\newline

We now show that all the eigenstates of $\hat{U}^{(n)}$ for ${\bf
w}^{(n)}= {\bf w}_{0}^{(n)}$ are ``crystalline''. By construction,
(\ref{qep}) are eigenstates of $\hat{D}_{j}=\hat{D}^{\prime }({\bf
R}_{j}^{\prime })$, $j=1,\ 2$, where $\hat{D}^{\prime }({\bf
R}_{j}^{\prime })$ are Weyl-Heisenberg translations in the $(u^{\prime
},\ v^{\prime })$ phase space by vectors ${\bf R}_{j}^{\prime }$
defining precisely the unit cell $T_{1}^{2}=(kB)^{-1}\cdot T^{2}$,
which is $k$ times smaller than $T^{2}$.  Thus, the Husimi density of
all the eigenstates of $\hat{U}^{\prime }$ for ${\bf w}^{\prime }={\bf
w}_{0}^{\prime }$ is exactly periodic on a lattice with unit cell
$T_{1}^{2}$. The eigenstates are then crystalline \cite{n}: each cell
contains $p$ of the $kp$ Husimi zeros in $T^{2}$. For $n>1$ and ${\bf
w}^{(n)}={\bf w}_{0}^{(n)}$, the $k^{n}p$ eigenstates of
$\hat{U}^{(n)}$ can be grouped into $n$ sets: the $l$th set,
$l=1,...,\ n$, consists of $k^{l-1}p(k+\delta _{l,1}-1)$ crystalline
eigenstates with unit cell $T_{n-l+1}^{2}$ and $k^{l-1}p$ Husimi zeros
in each cell. Unlike general crystalline eigenstates of the quantum
cat maps \cite{n}, the crystal features of all this complete basis of
eigenstates persist under an arbitrarily large perturbation
$\hat{U}_{1}^{(n)}$ in $\hat{U}^{(n)}$.
\newline

Another interesting result at fixed BCs concerns the expansion
(\ref{qes}) for the eigenstates (\ref{qep}) and can be easily derived
in the case when the vectors (\ref{sw}) form a square lattice in the
BZ, ${\bf w}=2\pi (r_{1},\ r_{2})/(gp)$, $r_{1},\ r_{2}=0,...,\ g-1$,
$g$ integer (see the conditions for this in note \cite{note1}). Since
$B=I/g$ \cite{note1}, the transformation (\ref{rt}) is simply ${\bf
z}=g{\bf z}^{\prime }$, and the eigenstates $\Psi _{b^{\prime },{\bf
w}_{0}^{\prime }}^{\prime }(u^{\prime }) $ of $\hat{U}^{\prime }$ can
be easily determined, without using (\ref{qep}), by just substituting
$u=gu^{\prime }$ in (\ref{qes}). After rearranging terms, we find that
$\Psi _{b^{\prime },{\bf w}_{0}^{\prime }}^{\prime }(u^{\prime })$ is
given by the expression in Eq. (\ref{qes}) with all the quantities
replaced by their primed counterparts and ${\bf w}^{\prime }={\bf
w}_{0}^{\prime }={\bf 0}$. For given $b$ and ${\bf w}$, an expansion
coefficient $\phi _{b^{\prime }(b,\ {\bf w})}(m^{\prime };\ {\bf w}
_{0}^{\prime })$, $m^{\prime }=0,...,\ p^{\prime }-1$, is nonzero only
if there exists an integer pair $(m,\ l)$, $m=0,...,\ p-1$, $l=0,...,\
g-1$, solving the Diophantine equation $pgl+gm+r_{1}=m^{\prime }$. The
solution is then unique and $\phi _{b^{\prime }}(m^{\prime };\ {\bf
w}_{0}^{\prime })=\exp (2\pi ilr_{2}/g)\phi _{b}(m;\ {\bf w})$,
associated with a {\em sparse} expansion. In particular, for $p=1$,
i.e., $\hbar ^{\prime }=2\pi /g^{2}$, we can choose $\phi _{b}(m;\
{\bf w})=1$, and the only nonzero coefficients are $\phi _{b^{\prime
}}(m^{\prime };\ {\bf w}_{0}^{\prime })=\exp (2\pi ilr_{2}/g)$ with
$m^{\prime }=gl+r_{1}$. This result can be obtained directly by
applying the methods in Refs. \cite{k,n}, where the general case of
$2\pi /\hbar =$ perfect square was studied in detail.\newline

In conclusion, we have shown that the quasienergy problem of general QAMs
for all BCs can be exactly reduced to a fixed-BCs problem. This reduction
is possible due to a distinctive feature of QAMs: For given $\hbar =2\pi
/p$, the $k$ allowed Bloch wave vectors ${\bf w}$ form a{\em \ lattice} in
the BZ, completely determined by the associated matrix $A$. Thus, this
reduction is not possible for the special torus maps with Tr$(A)=2$, e.g.,
the kicked rotor and the kicked Harper maps, since the number of BCs is
infinite for them. The reduction is implemented by a renormalization
transformation $ {\cal R}$ exhibiting nontrivial features: For Tr$(A)<-2$,
the quantum cat maps are always fixed points of ${\cal R}$, while for
Tr$(A)>2$ they are generally fixed points only of ${\cal R}$ accompanied
by time reversal. These features reflect the scaling invariance of the
quantum kernel (\ref{UA}). As a first application, we have shown that some
results at fixed BCs can be derived and understood in a simple and
transparent way in the general-BCs framework. In particular, one can
easily establish the existence of a complete fixed-BCs basis of
crystalline eigenstates in a classical limit for arbitrarily large
perturbations. The results in this Letter may be applied to study several
aspects of the general-BCs problem for QAMs using known results and
methods at fixed BCs. An important and unexplored issue is the sensitivity
of eigenstates to variations in the BCs and its relation to classical
phase-space structures, especially in a regime of mixed phase space (large
perturbations).\newline

{\bf Acknowledgments}\newline

The author would like to thank S. Nonnenmacher for helpful comments and
correspondence. Comments from J.P. Keating and Z. Rudnick are gratefully
acknowledged. This work was partially supported by the Israel Science
Foundation administered by the Israel Academy of Sciences and Humanities.


\begin{references}
\bibitem{note1} This case will occur only if $p{\bf y}/2$ in
(\ref{sw}) is an integer vector and the matrix $E$ can be chosen so
that $B=(A-I)^{-1}E=I/g $. This implies that $A=I+gE$. It is easy to
show that the latter relation is satisfied if and only if $A=sE^{2}$
with Tr$(E)=sg$, where $s=$ sgn[Tr$(A) $] $=$ $-$sgn[$\det (E)$]. One
can then easily find matrices $A$ for any $g$.
\end{references}
\end{document}